\documentclass[sigconf]{acmart}

\usepackage{graphicx}
\usepackage{subfiles}

\usepackage{todonotes}
\usepackage{booktabs}
\usepackage{multirow}
\usepackage{url}
\PassOptionsToPackage{hyphens}{url}\usepackage{hyperref}

\usepackage{dsfont}
\usepackage{amsmath}

\usepackage{caption}
\usepackage{subcaption}

\usepackage{xcolor}

\AtBeginDocument{%
  \providecommand\BibTeX{{%
    \normalfont B\kern-0.5em{\scshape i\kern-0.25em b}\kern-0.8em\TeX}}}

\copyrightyear{2024}
\acmYear{2024}
\setcopyright{rightsretained}
\acmConference[SIGIR '24]{Proceedings of the 47th International ACM SIGIR Conference on Research and Development in Information Retrieval}{July 14--18, 2024}{Washington, DC, USA}
\acmBooktitle{Proceedings of the 47th International ACM SIGIR Conference on Research and Development in Information Retrieval (SIGIR '24), July 14--18, 2024, Washington, DC, USA}
\acmDOI{10.1145/3626772.3657919}
\acmISBN{979-8-4007-0431-4/24/07}

\begin{document}

\title{Contextualization with SPLADE for High Recall Retrieval}

\author{Eugene Yang}
\affiliation{%
  \institution{Human Language Technology Center of Excellence, Johns Hopkins University}
  \city{Baltimore}
  \state{Maryland}
  \country{USA}
}
\email{eugene.yang@jhu.edu}

\begin{abstract}
High Recall Retrieval (HRR), such as eDiscovery and medical systematic review, is a search problem that optimizes the cost of retrieving most relevant documents in a given collection. 
Iterative approaches, such as iterative relevance feedback and uncertainty sampling, are shown to be effective under various operational scenarios. 
Despite neural models demonstrating success in other text-related tasks, linear models such as logistic regression, in general, are still more effective and efficient in HRR since the model is trained and retrieves documents from the same fixed collection.  
In this work, we leverage SPLADE, an efficient retrieval model that transforms documents into contextualized sparse vectors, for HRR. 
Our approach combines the best of both worlds, leveraging both the contextualization from pretrained language models and the efficiency of linear models. 
It reduces 10\% and 18\% of the review cost in two HRR evaluation collections under a one-phase review workflow with a target recall of 80\%. 
The experiment is implemented with TARexp and is available at \texttt{\url{https://github.com/eugene-yang/LSR-for-TAR}}.
\end{abstract}



\maketitle

\section{Introduction}

High recall retrieval (HRR), such as electronic discovery (eDiscovery)~\cite{baron2016perspectives, harty2017discovery} in litigation and medical systematic review~\cite{wallace2010semi}, considers the problem of minimizing the total cost of retrieving a large portion of relevant documents, i.e., high recall~\cite{cost-analysis}. 
These applications often pose a predefined recall target. 
Unlike recall-oriented search problems, achieving a high recall target is a requirement, and the goal is to minimize the cost of achieving such targets. 
Historically, documents were reviewed exhaustively~\cite{blair1985evaluation}. 
Search technologies, such as Boolean queries, were later applied to accelerate the process by filtering irrelevant documents~\cite{baron2007sedona, oard2013information, lewis2016defining}. 
HRR search processes recently are often iterative human-in-the-loop ones, named Technology-Assisted Review (TAR), that alternate between updating the search model and human experts reviewing additional documents~\cite{baron2016perspectives, oard2013information}. 

Prior works in HRR cost analysis have categorized TAR workflows into two main avenues: two-phase and one-phase workflows~\cite{cost-analysis}. A two-phase workflow, known as TAR 1.0, trains a classifier with active learning on training documents labeled iteratively. After the model is effective enough, we deploy the model to the remaining unreviewed documents. 
Documents predicted as relevant, especially in eDiscovery, are reviewed in a second phase, often by less expensive reviewers, to ensure the integrity of the process and verify the recall~\cite{cost-analysis, lewis2016defining}. 
A one-phase workflow, known as TAR 2.0, in contrast, continues to retrain the classifier and review documents until the recall target is met. 
In this case, the classifier does not stop training and continues to bring unreviewed documents that are most likely relevant to human attention~\cite{cormack2014evaluation, cormack2016engineering}. 
Despite serving different purposes, both workflows require a supervised learning model to iteratively prioritize documents for human review. 

While neural pretrained language models (PLMs), such as BERT~\cite{bert}, RoBERTa~\cite{roberta}, and T5~\cite{t5}, advance the effectiveness of numerous text-related tasks due to contextualization, linear models, such as logistic regression and SVM, are still very competitive for HRR~\cite{goldilocks}. 
Since both one and two-phase TAR workflows require updating the classifier with additional labels, retraining linear models is much more efficient than classification fine-tuning on the PLMs~\cite{goldilocks}. 
Besides efficiency, tailoring the PLM to the specific search collection in HRR, while critical, demands non-trivial hyperparameter tuning on the amount of language model fine-tuning before entering the iterative process but 
only moderately improves effectiveness, preventing PLMs from being incorporated into HRR. 
Recent work
in reproducing \citet{goldilocks} by \citet{goldilocks-repro} 
further confirms that the cause of this challenge is the domain mismatch between the search collection and the PLM. 

However, contextualization offered by PLM is still very valuable. 
In this work, instead of transforming a PLM as the underlying classifier, we employ it as a contextualized feature extractor. 
Inspired by SPLADE~\cite{splade}, an efficient learned-sparse retrieval model~\cite{thong-repro}, we propose using the masked language modeling ability in PLMs to encode documents as sparse vectors and use them as features for linear models during TAR. 
This combines the best of both worlds -- the contextualization from the PLMs and the efficiency from the linear classification models. 

The contribution of this work is three-fold:
(1) an effective sparse classification model for HRR,
(2) comprehensive experimentation with two workflows on two HRR evaluation collections, and 
(3) an ablation study on the impact of the choice of PLM.

\section{Background}

Various linear models have been studied in the context of HRR, which differs from the typical text classification or ad hoc retrieval, where models are trained to generalize to unseen text or queries. 
In contrast, HRR problem is transductive, meaning that the model is trained and used for inference on the same document collection~\cite{cost-analysis}. 
Models, such as logistic regression, SVM, and KNN, are generally equally effective with appropriate text features~\cite{yang2017effectiveness}. 
On the other hand, sampling strategies have a larger impact on the overall effectiveness of the TAR process. 
Iterative relevance feedback, advertised as Continuous Active Learning\textsuperscript{TM} in eDiscovery~\cite{cormack2014evaluation, cormack2016engineering}, is shown to be more efficient in achieving the plateau performance in retrieving relevant documents in a one-phase workflow than uncertainty sampling~\cite{lewis1994sequential, lewis1994heterogeneous} and simple random sampling. 

Further investigations suggest that the choice of workflows and sampling strategies depends on the cost structure~\cite{cost-analysis}. 
With no review cost difference between the two phases, 
one-phase iterative relevance feedback 
yields the lowest total reviewing cost for the same underlying classification model. 
However, in eDiscovery, documents for training the classifier may require more careful review or by senior personnel, implying a higher cost in the first phase~\cite{cost-analysis}. 
In this case, a two-phase workflow with uncertainty sampling yields a lower overall cost than other combinations. 
In this work, we experiment with both one-phase workflow with iterative relevance feedback and two-phase workflow with uncertainty sampling to demonstrate the robustness of the proposed method. 

Since the introduction of the transformer models, baseline effectiveness of almost all text-related tasks has been raised by simply adapting pretrained language models~\cite{bert}. 
In ad hoc retrieval, cross-encoders~\cite{monobert, monot5}, which concatenate the query and document together as the input for text pair classification, demonstrate strong effectiveness but are inefficient at search time.
Neural text encoders trained with representation learning~\cite{cocondenser, splade} spawn the dual-encoders~\cite{colbert, dpr, splade}, which encode documents into contextualized vectors at indexing time and only compute the lightweight similarity between the query and document vectors at search time. 

Dual-encoders fall into two categories: dense, such as DPR~\cite{dpr} and ColBERT~\cite{colbert}, where text are encoded as one or more dense vectors, and learned-sparse retrievers (LSR)~\cite{thong-repro}, such as EPIC~\cite{macavaney2020expansion} and SPLADE~\cite{splade}, where text are transformed into contextualized sparse vectors.
Despite recent engineering efforts on dense retrievers~\cite{faiss, plaid}, the integration of LSR with extremely efficient retrieval architectures, such as Apache Lucene and PISA~\cite{pisa}, still demonstrates lower query latency while offering competitive effectiveness. 

Prior work has experimented with incorporating cross-encoders with HRR~\cite{goldilocks, goldilocks-repro}.
However, the cross-encoders are barely more effective than baseline logistic regressions after careful language model fine-tuning. 
Since the model needs to perform inference on all unreviewed documents in every iteration, the inefficiency of cross-encoders eliminates them from any practical use. 
Note that \citet{goldilocks} only experimented with 20\% of RCV1 when applying cross-encoder due to GPU limitations, showing how inefficient the process is. 
In this work, we aim to improve both effectiveness and efficiency when incorporating PLMs with HRR through LSR. 

In the next section, we describe our effort in leveraging LSR as a contextualized sparse encoder for HRR.

\section{Contextualized Text Features with Learned-Sparse Retrieval}

Before entering the iterative TAR process, documents are tokenized and encoded as features in preparation for supervised learning. 
In prior works, documents are encoded with statistical weights such as TF-IDF or BM25-saturated weights~\cite{yang2019text, yang2019regularization}. 
To extract contextualized textual features, we leverage the masked language model (MLM) head trained with PLMs, such as BERT and RoBERTa~\cite{splade}. 
However, since plain MLM heads are trained to reconstruct original text with surrounding context~\cite{bert}, the prediction over the vocabulary space is not necessarily sparse, which is, however, desired for efficiency. 
Additional training on retrieval training data, such as MS MARCO~\cite{msmarco}, can further adjust the prediction vector to be sparse and tailored for capturing relevancy between texts~\cite{thong-repro}. 

To encode a document $d_i$ in the search collection, we apply MLM $F(t) \rightarrow \mathbb{R}^m$ on each document token $t_{ij}$ where $m$ is the size of the vocabulary space, aggregate the predictions from all $t_{i\cdot}$ tokens by taking the maximum score for each predicted token, and select the top $s \le m$ predicted tokens with the highest scores. 
Therefore, the document-token matrix of the entire collection is a sparse matrix $M \in \mathbb{R}^{n\times m}$ where $n$ is the number of documents. 

We select a handful of seed documents at the beginning of the TAR iterative process. 
In practice, they are usually selected based on exploratory searches from the users to gain an initial understanding of the search task~\cite{oard2010evaluation}.
For experimentation, in this work, we randomly select one relevant and one non-relevant document to start. 
In each TAR iteration, a logistic regression model is retrained with the reviewed documents, unreviewed documents using the updated models, and a batch of documents is sampled for human review. 
The sample can be based on relevance feedback, i.e., unreviewed documents with the highest scores, uncertainty sampling, i.e., unreviewed documents with a predicted 0.5 probability of being relevant, or random sampling.

\section{Experiments}

We evaluate the proposed method on two HRR evaluation collections: RCV1-v2~\cite{rcv1} and Jeb Bush~\cite{totalrecall2015, totalrecall2016}. 
RCV1-v2 contains 804,414 news articles from Reuters with 658 hierarchical categories, including 102 topics, 324 industries, and 232 regions. 
Each document is categorized into one or more categories by journalists from Reuters. 
We use the curated 45 categories from \citet{cost-analysis} with three levels of difficulties and prevalence. 
Jeb Bush collection contains 274,124 emails (after deduplication with MD5 hashes of the content) from Jeb Bush during his tenure as the Governor of Florida in the United States. 
44 topics (referred to as ``categories'' in the rest of the paper for consistency) were created during TREC Legal Track in 2015 and 2016~\cite{totalrecall2015, totalrecall2016}. 
Recall target is set as 80\% in this study. 

Our TAR workflow is implemented with TARexp~\cite{tarexp}\footnote{\url{https://github.com/eugene-yang/tarexp}}, a Python framework for running TAR experiments. 
We conducted experiments on both one-phase workflows with iterative relevance feedback under a uniform cost structure (no cost difference in reviewing any documents) and two-phase workflows with uncertainty sampling under an expensive training cost structure (training documents are ten times more expensive to review). 
The two workflows are shown to be optimal under each cost structure~\cite{cost-analysis}. 
One relevant and one non-relevant documents are randomly selected as the seed documents for TAR.
We execute each TAR workflow on the same category with ten different random seeds, resulting in 890 (450 in RCV1 and 440 in Jeb Bush) runs for each experimented workflow. 
Experiments on the same category all start with the same ten seed sets. 
In each TAR iteration, we sample 200 unreviewed documents using either relevance feedback or uncertainty sampling and reveal their true labels as if they were annotated by a human. A logistic regression is retrained with the updated reviewed document set.
We employ an oracle stopping rule used in \citet{cost-analysis} to stop the workflow when the true recall meets the target.
Determining when to stop is a rich line of work~\cite{cikm2021, li2020stop, stevenson2023stopping, yang2021heuristic, lewis2021certifying} beyond the scope of this work.

We use the publicly available SPLADE~\cite{formal2022distillation} on HuggingFace trained with knowledge distillation and CoCondenser~\cite{cocondenser}, which is a strong LSR model.\footnote{\url{https://huggingface.co/naver/splade-cocondenser-ensembledistil}}
Since the model is fine-tuned from BERT~\cite{bert}, the size of the vocabulary space is 30522.
We select the top 10\% from the predicted tokens, which results in at most 3052 tokens for each document. 
Note that since the SPLADE model was trained with sparsity regularization, each document may have fewer than the maximum number of tokens. 
In fact, each document has, on average, 194.85 and 194.24 predicted tokens for RCV1 and Jeb Bush, respectively. 
We compare the proposed method with encoding documents with BM25-saturated term frequency, which is shown to be more effective than TF-IDF~\cite{yang2019text}. 

We also experiment with combining BM25 and SPLADE encoders during TAR. 
Specifically, at each TAR iteration, we train two logistic regression models using BM25 and SPLADE text features separately. 
For each unreviewed document, we average the predicted probabilities from the two models and pass them to the sampling strategy for the rest of the TAR process. 

\begin{table}[t]
\caption{Cost reduction on RCV1 and Jeb Bush using the two workflows compared to BM25. }\label{tab:main-two-coll}
    \centering
    \begin{tabular}{l|cc|cc}
\toprule
{} & \multicolumn{2}{c|}{One-Phase w/ Rel} & \multicolumn{2}{c}{Two-Phase w/ Unc} \\
{} & \multicolumn{2}{c|}{Uniform Cost} & \multicolumn{2}{c}{Expensive Training} \\
\midrule
              & Jeb Bush &    RCV1 &Jeb Bush &    RCV1 \\
\midrule
BM25-Saturated &  1.0000 &  1.0000 &  1.0000 &  1.0000 \\
SPLADE         &  0.9467 &  1.0122 &  0.9292 &  0.7912 \\
\midrule
BM25 + SPLADE  &  \textbf{0.8260} &  \textbf{0.8957} &  \textbf{0.7520} &  \textbf{0.7265} \\
\bottomrule
\end{tabular}
    
\end{table}

\begin{figure*}[t]
    \centering
    \includegraphics[width=\textwidth]{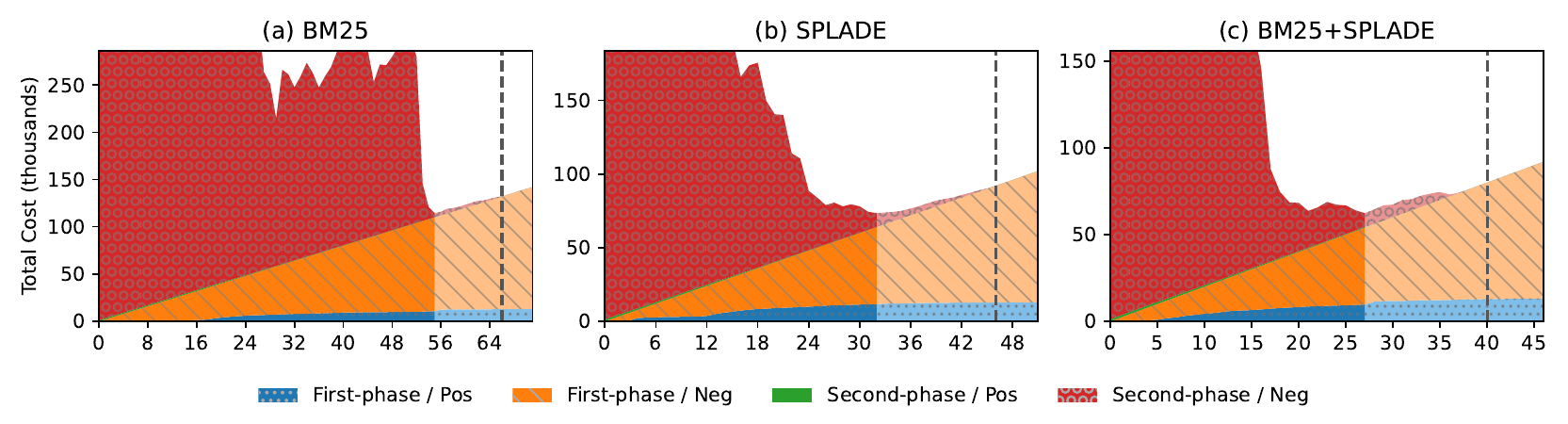}
    \vspace{-2em}
    \caption{Cost dynamic graphs on category \texttt{COMMERCIAL VEHICLES}(I35102), which is a hard and rare category in RCV, using a two-phase workflow with uncertainty sampling under an expensive training cost structure. The gray dashed vertical line indicates where the iteration requires no second phase review, i.e., first phase review already reaches 80\% recall. }
    \label{fig:example-2p-workflow}
\end{figure*}

We report the relative optimal total review cost~\cite{cost-analysis} compared to the baseline BM25 model as the effectiveness metric. 
Values larger than 1.0 indicate higher cost and smaller than 1.0 indicate cost reduction compared to using BM25. 
For one-phase workflows with a uniform cost structure, the total review cost is defined as the number of documents (relevant and non-relevant ones combined) before reaching 80\% recall.
For two-phase workflows with training documents being ten times more expensive than the rest, the total cost is defined as ten times the number of documents being reviewed during the first phase plus the number of additional documents (sorted by predicted scores) required to be reviewed in the second phase to reach 80\% recall. 
Since in a two-phase workflow, any iteration can reach 80\% recall as long as one can afford to go deep in the rank list, we report the optimal total cost over the iteration, i.e., the lowest total review cost for a run can possibly get. Figure~\ref{fig:example-2p-workflow} contains three example runs with two-phase workflow, which we will discuss in detail in the next section. 
\section{Results and Analysis}

Table~\ref{tab:main-two-coll} summarizes experiments on the two workflows.
When using SPLADE as the document feature extractor, it provides a moderate effectiveness improvement over BM25.
Such contextualized features are more helpful in the two-phase workflow with uncertainty sampling when the primary goal is to produce the most effective classifier sooner.
These features provide a better representation of each document, resulting in a 10\% to 20\% cost reduction. 
Since the contextualization is essentially co-occurrence estimated on large corpora, these features import external language knowledge to the task, which is lacking when using BM25. 
In contrast, since iterative relevance feedback in a one-phase workflow continuously feeds the top-ranked documents for human review, it does not need to learn all aspects of relevancy at once as long as the proximity of the reviewed relevant documents still supplies enough additional relevant documents for human review. 
Therefore, the benefit of better textual features provided by SPLADE is less apparent in the one-phase workflow case, with only 5\% cost reduction on Jeb Bush. 

By combining both BM25 and SPLADE during active learning, we further improve the effectiveness on both one and two-phase workflows. 
There are still valuable complementary signals in the surface form of the text, which may be lost after the contextualization of the transformer model. 
Combining top candidates from BM25 and SPLADE, we further improve the effectiveness on both collections and both workflows, resulting in 10\% to 27\% cost reduction.

\subsection{Other Language Models}
\begin{table}[]
\caption{Ablation with different language model. }\label{tab:diff-matrix-one-phase}
    \centering

\begin{tabular}{l|cc}
\toprule
Language Model                &Jeb Bush &    RCV1 \\
\midrule
BM25-Saturated                &  1.0000 &  1.0000 \\
\midrule
BERT Large                    &  1.8839 &  1.1954 \\
BERT Large >> MLM 2k Steps    &  1.9486 &  1.1932 \\
BERT Large >> MLM 100k Steps  &  1.7022 &  1.1807 \\
\midrule
MLM From Scratch 200k Steps   &  1.8050 &  1.9909 \\
MLM From Scratch 400k Steps   &  1.7623 &  1.8109 \\
\midrule
SPLADE                        &  0.9467 &  1.0122 \\
BM25 + SPLADE                 &  \textbf{0.8260} &  \textbf{0.8957} \\
\bottomrule
\end{tabular}

\end{table}

Table~\ref{tab:diff-matrix-one-phase} presents experiments with one-phase workflow with iterative relevance feedback using five models, including four that are fine-tuned on the search collection without retrieval objectives. 
We observe that the out-of-box BERT Large model incurs a 19\% cost increment on RCV1 and a large 88\% on Jeb Bush. 
This difference in cost aligns with prior findings on domain mismatch when applying cross-encoders to TAR~\cite{goldilocks, goldilocks-repro}. 
Additional language model fine-tuning with MLM on the search collection mitigates such mismatches, dropping the cost increment from 88\% to 70\% on Jeb Bush after 100k steps. However, it is still far worse than just using BM25. 

Language models trained from scratch using the search collection do not lead to better effectiveness either. 
Without training on large external corpora, the MLM encoder is still worse than BM25.
We conclude that retrieval fine-tuning on SPLADE is more valuable than tuning the language to the specific search collection. 

\subsection{Category Analysis on RCV1}

\begin{table}[]
\caption{RCV1 category groups with one-phase workflow. }\label{tab:rcv1-one-phase-break-down}
    
\centering
\begin{tabular}{ll|ccc}
\toprule
Difficulty & Prevalence &    BERT &  SPLADE &  BM25+SPLADE \\
\midrule
\multirow{3}{*}{Hard}
& Rare   &          1.0566 &      0.8885 & \textbf{0.5566} \\
& Medium &          1.1166 &      0.8527 & \textbf{0.8061} \\
& Common &          1.2552 &      0.9920 & \textbf{0.8536} \\
\midrule
\multirow{3}{*}{Medium}
& Rare   &          1.3964 &      1.1877 & \textbf{0.8876} \\
& Medium &          1.4026 &      1.7508 & \textbf{1.1650} \\
& Common &          1.2038 &      0.9409 & \textbf{0.9128} \\
\midrule
\multirow{3}{*}{Easy}
& Rare   &          1.2307 &\textbf{0.8806} &      0.9081 \\
& Medium &          1.0141 &\textbf{0.9783} &      0.9797  \\
& Common &          1.0129 &      0.9904 & \textbf{0.9893} \\
\bottomrule
\end{tabular}

\vspace{-1em}
\end{table}
We take a closer look at the one-phase workflow on the RCV1 categories by breaking them into nine groups in Table~\ref{tab:rcv1-one-phase-break-down}. 
Both BERT Large and SPLADE demonstrate better effectiveness in hard categories compared to other groups, with SPLADE being more effective in almost all groups. 
Interestingly, additional retrieval fine-tuning received by SPLADE reduces the effectiveness on categories with medium difficulty and prevalence, which are 
\texttt{FOOD, DRINK AND TOBACCO PROCESSING} (I41000), 
\texttt{IRON AND STEEL} (I22100), 
(financial) \texttt{RESERVES} (E513), 
\texttt{AEROSPACE} (I36400), and
\texttt{TOBACCO} (I42900). 
It is likely an artifact of training SPLADE with MS MARCO, where content related to certain topics may be underrepresented. 

However, combining SPLADE with BM25 almost always improves effectiveness.
Especially on rare and hard categories, which are ones for which statistical features such as BM25 struggle the most, BM25 with SPLADE provides an additional 33\% cost reduction from only using SPLADE. 
Categories that are easy (which were measured by logistic regression using statistical features~\cite{cost-analysis}) and rare do not benefit from additional signals with contextualized features, which is expected since BM25 features are already well-equipped for retrieval in such tasks. 
However, including SPLADE features does not, or only slightly, degrades effectiveness.

\subsection{Cost Dynamics with SPLADE features}

Figure~\ref{fig:example-2p-workflow} are three cost dynamic graphs with two-phase workflows using uncertainty sampling. 
This analytical tool, which was introduced by \citet{cost-analysis}, 
separates the total cost at each iteration into four sectors. 
Specifically, the amount of the second phase cost is calculated based on the depth of the ranked list, sorted in decreasing order by the predicted probability by the classifier trained at that iteration. 
Figure~\ref{fig:example-2p-workflow}a shows that BM25 struggled for 55 iterations without significantly reducing the number of second phase non-relevant documents needed to review before reaching the recall target, which is a clear indicator of failing to improve the classifier fast enough. 
Workflow using SPLADE (Figure~\ref{fig:example-2p-workflow}b), in contrast, continues to improve the classifier, 
resulting in far fewer iterations (31) before reaching 80\% recall. 
Combining BM25 and SPLADE (Figure~\ref{fig:example-2p-workflow}c), similar to only SPLADE, allows the classifier to continuously improve over the course of TAR but in a faster pace. 

\section{Summary}

In this work, we investigate using SPLADE as a contextualized sparse document encoder for TAR. 
Prior work that combines cross-encoder with TAR not only is extremely inefficient, but also fails to provide satisfying effectiveness improvement. 
Experiments on two evaluation collections, our proposed approach on combining contextualized features from SPLADE and statistical features using BM25-saturated weights with logistic regression brings the best of both worlds, providing a large cost reduction in the TAR process.

\begin{acks}
The author thanks Jeremy Pickens and James Mayfield for their constructive feedback and proofreading. 
\end{acks}

\bibliographystyle{ACM-Reference-Format}
\bibliography{sample-base}

\end{document}